\documentclass{article}
\usepackage{graphicx}
\usepackage{amsmath}

\input{tcilatex}

\begin{document}

\begin{center}
{\LARGE LHC Based ep Colliders: e-linac versus e-ring}

\bigskip

Y. Islamzade$^{1)}$,{\large \ }H. Karadeniz$^{1,2)}$, S. Sultansoy$^{1,3)}$
\end{center}

$^{1)}$Dept. of Physics, Faculty of Arts and Sciences, Gazi University,
06500 Teknikokullar, Ankara, Turkey.

$^{2)}$Ankara Nuclear Research and Training Center, 06100 Besevler, Ankara,
Turkey.

$^{3)}$Institute of Physics, Academy of Sciences, H. Cavid Ave. 33, Baku,
Azerbaijan.

\bigskip

\bigskip

\begin{center}
\textbf{Abstract}
\end{center}

Luminosity estimation for Linac$\ast $LHC based ep collider is evaluated and
compared to the suggested ''LEP''$\ast $LHC type collider.

\bigskip

\textbf{I. Introduction}\bigskip

The importance of lepton-hadron collisions is accepted by the high energy
physics community [1]. There are two ways of colliding a ring protons with
electrons: accelerating electrons by a linac or in a ring. Both methods have
advantages and disadvantages of their own, but an overall evaluation shows
e-linacs to be preferable to e-rings at TeV c.m. energy scale.

Low current I which constrains luminocity L and the hourglass effect are the
main disadvantages of linac-ring type machines. On the other hand, there is
no synchrotron radiation in linacs and the spin manipulation is much more
easier.

In our previous study [2] we show Linac$\ast $VLHC option to be preferable
compared to ring$\ast $VLHC option [3] for VLHC based ep collider. In this
note, we perform similar analysis for LHC based ep colliders.

In an article by E. Keil [4], the ''LEP''$\ast $LHC collider parameters are
estimated (see Table 1). We suggest a Linac$\ast $LHC combination with
almost better parameters and put forward some objections:

i) Linac$\ast $LHC with L$=2.0$x$10^{32}$ cm$^{-2}$s$^{-1}$ will give
opportunity to additional $\gamma $p and $\gamma $A options.

ii) Constructing a linac of less than 3 kms length is more economical than
to construct an electron ring in the LHC tunnel. Moreover, the last version
will lead to some technical problems.

iii) For Linac$\ast $LHC based colliders, energy can be increased by
extending linacs.\bigskip

\textbf{II. Linac}$\ast $\textbf{LHC based ep colliders}\bigskip

Luminosity of Linac$\ast $LHC ep collider is estimated using luminosity
expression for the TESLA$\ast $HERA ep collider [5] with TESLA as a
prototype for other linacs (e.g., CLIC, JLC/NLC):

\begin{equation*}
\text{L}=4.8\text{x}10^{30}\text{cm}^{-2}\text{s}^{-1}\frac{\text{N}_{\text{P%
}}}{10^{11}}\frac{10^{-6}\text{m}}{\in _{\text{P}}}\frac{\gamma _{p}}{1066}%
\frac{10\text{cm}}{\beta ^{\ast }}\frac{\text{P}_{\text{e}}}{22.6\text{MW}}%
\frac{250\text{GeV}}{\text{E}_{\text{e}}}\text{ \ \ \ \ \ \ \ \ \ \ \ \ (1)}
\end{equation*}
where subscripts e and p denote the linac electron beam and the proton ring
beam parameters, respectively. Taking N$_{\text{P}}=$10$^{11}$, $\in _{\text{%
P}}=10^{-6}$m, $\gamma _{p}=$7500, $\beta ^{\ast }=0.1$ m the formula
becomes:

\bigskip

\begin{equation*}
\text{L}=3.4\text{x}10^{31}\text{cm}^{-2}\text{s}^{-1}\frac{\text{P}_{\text{e%
}}}{22.6\text{MW}}\frac{250\text{GeV}}{\text{E}_{\text{e}}}\text{ \ \ \ \ \
\ \ \ \ \ \ \ \ \ \ \ \ \ \ \ \ \ \ \ \ \ \ \ \ \ \ \ (2)}
\end{equation*}

\bigskip

For a generic and rough evaluation of luminosity for all the three linacs
(CLIC, JLC/NLC, TESLA) we take P$_{\text{e}}$ and E$_{\text{e}}$ to be the
same. The synchrotron radiation power in the e-ring is 34.5 MW [4]. This
amount of power must be supplied to beams by the RF system. Since there is
no synchrotron radiation in e-linacs, inserting this value into (2) and
taking E$_{\text{e}}=67.3$ GeV (the same as Keil's e-beam), luminosity
becomes:

\bigskip

\begin{equation*}
\text{L}=2.0\text{x}10^{32}\text{cm}^{-2}\text{s}^{-1}
\end{equation*}

\bigskip

The Linac$\ast $LHC luminosity can be further increased by applying the
''dynamic focusing'' method developed by Brinkmann and Dohlus which ''...
would allow to increase the luminosity by at least a factor of three'' [6].
In addition, proton phase density (N$_{\text{P}}$/$\in _{\text{P}}$) can be
essentially higher than 10$^{17}$m$^{-1}$ since IBS in the main ring is not
crucial at the LHC because of the large value of $\gamma _{p}$. To use this
advantage, proton beam must be cooled before injection into the main ring.
We expect an increase of luminosity at least by a factor of three in this
way. Both of these methods applied to Linac$\ast $LHC will be presented in
our next study [7].\bigskip

\textbf{III. Conclusion}\bigskip

Of the two methods of colliding electrons with protons, Linac$\ast $LHC
combination seems to be advantageous. A very rough calculations yield
luminosity of L=$2.0$x$10^{32}$cm$^{-2}$s$^{-1}$ which is greater than
Keil's L=$1.2$x$10^{32}$cm$^{-2}$s$^{-1}$ value (It is worth to remind that $%
\sqrt{\text{s}}$=1.37 TeV in both options). Moreover, with applying
''dynamic focusing'' and proton beam cooling the luminosity value of order
of $10^{33}$cm$^{-2}$s$^{-1}$ seems to be quite realistic for our option.

The other parameters depend on the particular choise of a linac. For
example, the lenght of TESLA-type linac will be $\approx 3$ km ($\approx 0.4$
km for CLIC and $\approx 1.4$ km for JLC/NLC). The advances in accelerating
systems will increase the accelerating gradient (see, for example ref. [8])
and decrease linacs lenght.

Linac$\ast $LHC combination, thus, is cost effective, flexible and offers
additional $\gamma $p and $\gamma $A options. The E$_{e}=$67.3 GeV energy
linac can be considered as the first step to the full Linac$\ast $LHC
machine [9] with E$_{e}=1$ TeV and $\sqrt{\text{s}}=5.29$ TeV.

\bigskip

\textbf{Acknowledgements}

\bigskip

This work is supported in part by Turkish Plannig Organization (DPT) under
the Grant No 2002K120250.

\bigskip

\textbf{References}

\bigskip

[1] S. Sultansoy, Turk. J. of Phys. 22, 575 (1998).

[2] Y. Islamzade, H. Karadeniz and S. Sultansoy, hep-ex/0204034 (2002).

[3] J. Norem and T. Sen, FERMILAB-PUB-99/347 (1999).

[4] E. Keil, LHC Project Report 93, CERN (1997).\ \ \ \ \ 

[5] TESLA Technical Design Report, DESY 2001-011.

[6] R. Brinkmann and M. Dohlus, DESY-M-95-11 (1995).

[7] Y. Islamzade, H. Karadeniz and S. Sultansoy, in preparation.

[8] CERN COURIER, V42, N5, 2002.

[9] O. Yavas, A.K. Ciftci and S. Sultansoy, Proc. of the 7 th European
Accelerator Conference (EPAC2000), p.391; hep-ex/0004016.

\begin{center}
\bigskip

\bigskip

\bigskip

\bigskip

\bigskip

\textbf{Table 1.} ''LEP''$\ast $LHC e and p beam parameters.\bigskip 

\begin{tabular}{|l|l|l|}
\hline
& e & p \\ \hline
Energy, E [GeV] & 67.3 & 7000 \\ \hline
Bunch population, N & 6.4x10$^{10}$ & 10$^{11}$ \\ \hline
Beam current, I [mA] & 59 & 92 \\ \hline
Number of bunches & 508 & 508 \\ \hline
Radiation power, W(MW) & 34.5 & - \\ \hline
\end{tabular}
\end{center}

\end{document}